\documentclass[fleqn,a4paper]{article}

\usepackage[utf8]{inputenc} 
\usepackage[T1]{fontenc} 
\usepackage{amsmath,amssymb,latexsym,graphicx}
\usepackage{newtxtext,newtxmath} 
\usepackage[dvipsnames]{xcolor}
\usepackage[
  colorlinks=true,
  urlcolor=MidnightBlue,
  citecolor=MidnightBlue,
  filecolor=MidnightBlue,
  linkcolor=MidnightBlue
]{hyperref} 
\usepackage[
  style=phys,
  eprint=true,
  maxnames=100
]{biblatex}
\usepackage{anyfontsize}
\usepackage{authblk}

\begin{document}

\title{
  Log-correlated color in Monet's paintings
}
\author{Jaron Kent-Dobias}
\affil{Laboratoire de Physique de l'Ecole Normale Supérieure, Paris, France}
\maketitle
\begin{abstract}
  We describe evidence for logarithmic correlations within the paintings of
  Claude Monet.
\end{abstract}

Logarithmic correlations appear in a variety of physical models---from
statistical mechanics to quantum field theory to turbulence---and play an
important role in several fields of pure and applied mathematics---from random
matrix theory to number theory to finance \cite{Berestycki_2015_Introduction,
Carpentier_2001_Glass, Duplantier_2017_Log-correlated, Fyodorov_2016_Moments,
Fyodorov_2014_Freezing, Fyodorov_2016_Fractional, Kistler_2014_Derridas,
Rhodes_2014_Gauussian, Bailey_2022_Maxima, Cao_2018_Log-correlated,
Fyodorov_2020_Statistics, Fyodorov_2012_Freezing, Sheffield_2007_Gaussian}.
Here, we describe evidence for logarithmic correlations in the world of art.

\begin{figure}
  \centering
  \includegraphics[width=0.8\textwidth]{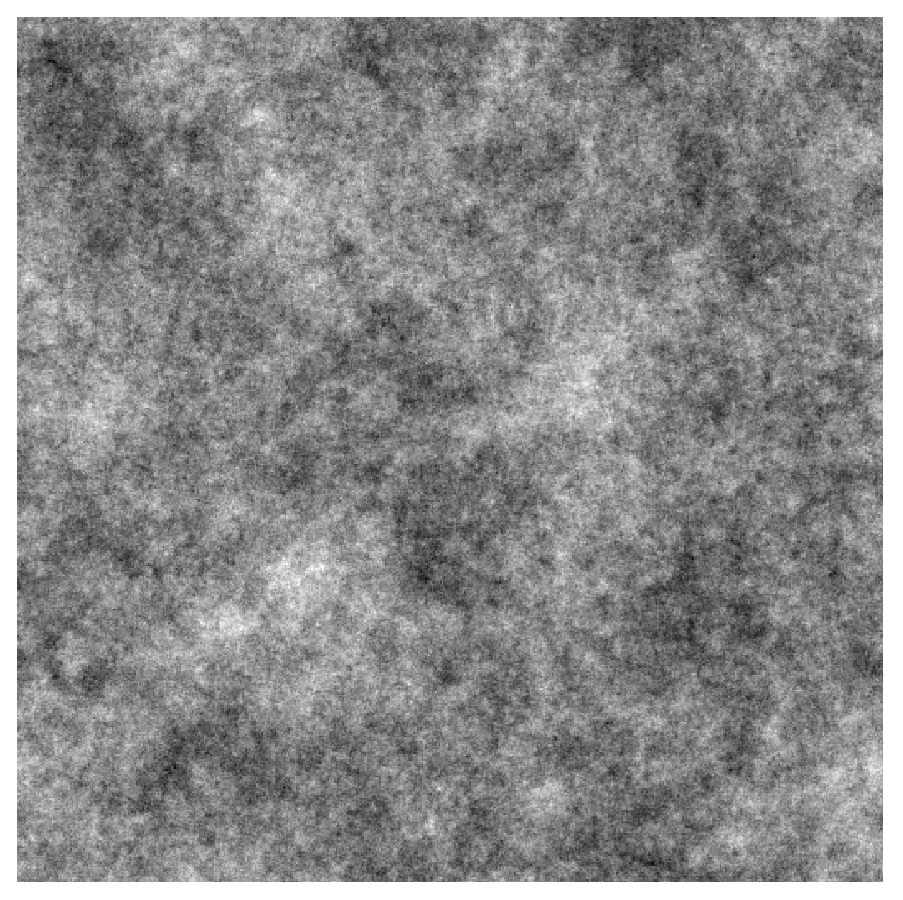}
  \caption{
    A typical equilibrium configuration of the 2D Gaussian free field on
    $512\times512$ square lattice at $T=1$.
  } \label{fig:gff}
\end{figure}

\begin{figure}
  \centering
  \includegraphics[width=0.9\textwidth]{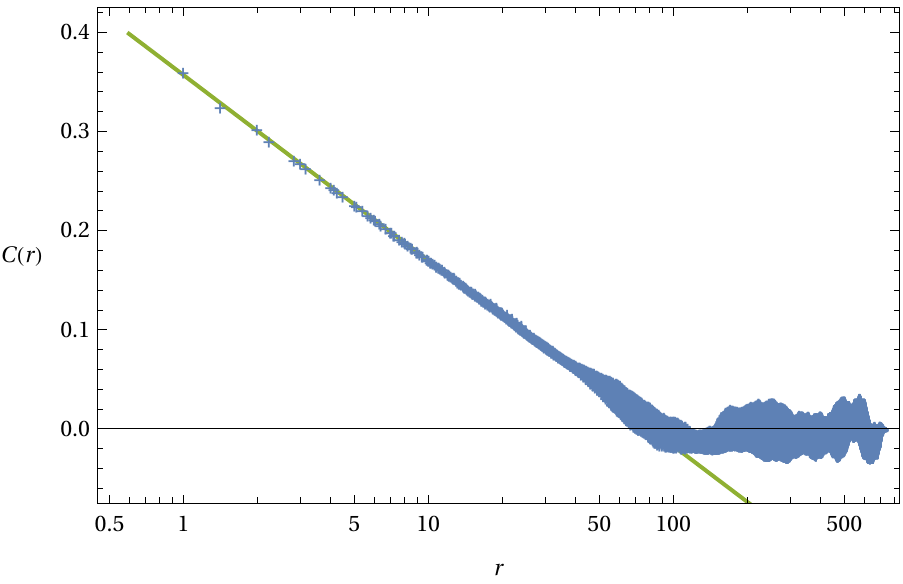}
  \caption{
    The empirical correlation function of the 2D Gaussian free field resulting
    from the snapshot in Fig.~\ref{fig:gff}. The green line shows a fit of
    \eqref{eq:corr}. Logarithms appear linear on a log--linear plot.
  } \label{fig:gff_corr}
\end{figure}

The two-dimensional Gaussian free field is a simple physical model with
logarithmic correlations. On a lattice, the model consists of harmonic springs
connecting the height of neighboring lattice sites. A typical equilibrium
configuration of this model is shown in Fig.~\ref{fig:gff}. The correlation
function takes the form
\begin{equation} \label{eq:corr}
  C(r)=\big\langle\big(h(\mathbf x)-\langle h\rangle\big)\big(h(\mathbf x')-\langle h\rangle\big)\big\rangle_{\|\mathbf x-\mathbf x'\|=r}\propto\log\frac{r_0}r
\end{equation}
The logarithmic correlations can be seen in averages taken over a single
typical configuration. Fig.~\ref{fig:gff_corr} shows the empirical correlation
function derived from averaging over the snapshot in Fig.~\ref{fig:gff}.
Logarithmic correlations result in a fractal geometry of fluctuations, along
with other rich properties that have been researched extensively
\cite{Bacry_2013_Log-normal, Duchon_2010_Forecasting}.

\begin{figure}
  \centering
  \includegraphics[width=0.8\textwidth]{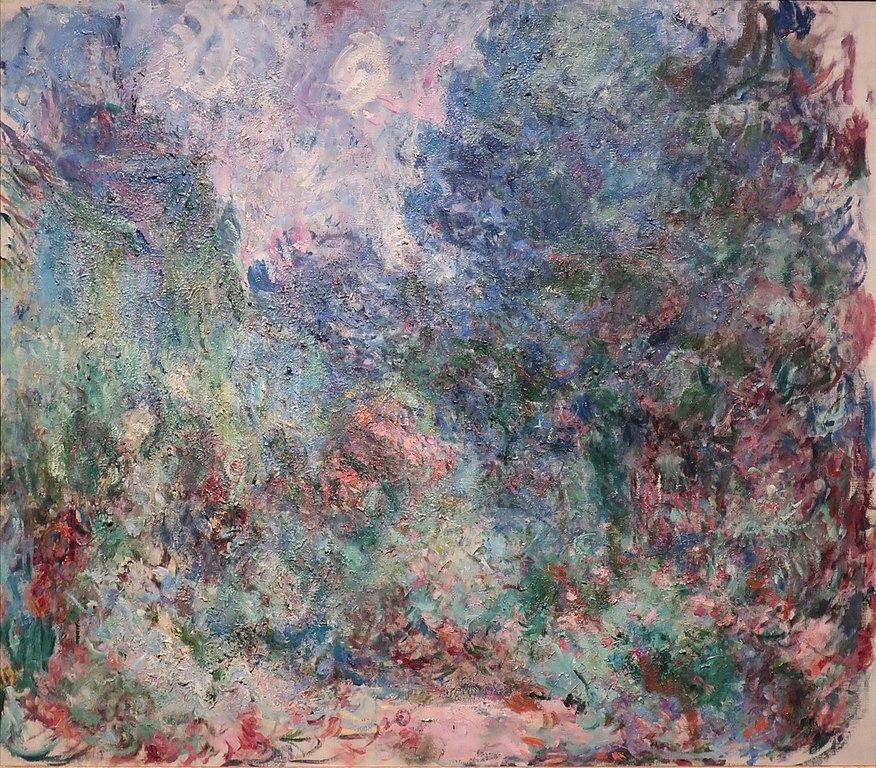}
  \caption{
    \href{https://commons.wikimedia.org/wiki/File:The_Artist\%27s_House_Seen_from_the_Rose_Garden_by_Claude_Monet_5103.JPG}{\textit{The artist's house seen from the rose garden}, by Claude Monet (1922--1924)}.
  } \label{fig:rose_garden}
\end{figure}

Examining Fig.~\ref{fig:gff}, one might notice a superficial similarity with
impressionist art. Consider the painting in Fig.~\ref{fig:rose_garden},
\textit{The artist's house seen from the rose garden} by Claude Monet. This
painting has fluctuations of color that strike the authors as qualitatively
alike to that of the Gaussian free field.  In the following, we argue that
these fluctuations are also \emph{quantitatively} alike.

\begin{figure}
  \centering
  \includegraphics[width=0.9\textwidth]{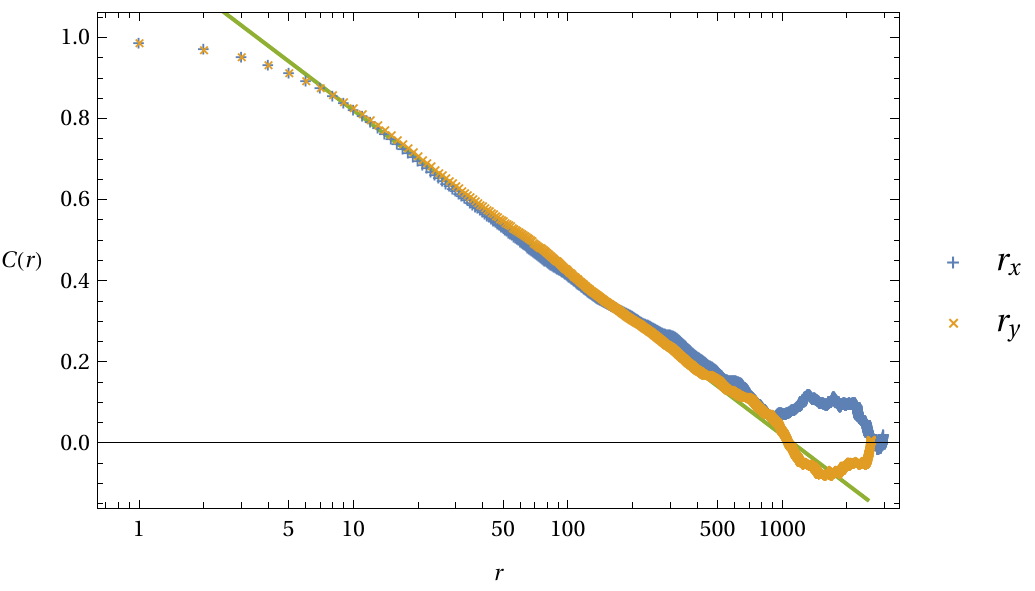}
  \caption{
    The correlation function along the horizontal ($\mathbf r=r_x\hat{\mathbf x}$)
    and vertical ($\mathbf r=r_y\hat{\mathbf y}$) for the hue field of the
    painting in Fig.~\ref{fig:rose_garden}. The line shows a fit of a
    logarithmic decay.
  } \label{fig:rose_corr}
\end{figure}

To examine correlations within the painting in Fig.~\ref{fig:rose_garden}, we
first convert its pixel color encoding to hue, saturation, and brightness
(HSB). Hue, in particular, can be represented as a point on a periodic color
line. We extract the hue from the image and encode its value as a vector on the
unit circle. Encoded in this way, the hue of the image is equivalent to a
state of the \textsc{xy} model on a square lattice. This allows us to define the
correlation function
\begin{equation}
  C(\mathbf r)=\big\langle\mathbf h(\mathbf x)\cdot\mathbf h(\mathbf x')\big\rangle_{\mathbf r=|\mathbf x-\mathbf x'|}
\end{equation}
The resulting correlation function is plotted in Fig.~\ref{fig:rose_corr},
along with a logarithmic fit. Remarkably, the correlations in the hue are
described extremely well by a logarithm with a short-distance cutoff. The
horizontal and vertical directions in this painting have nearly identical
correlations.

\begin{figure}
  \centering
  \includegraphics[width=0.8\textwidth]{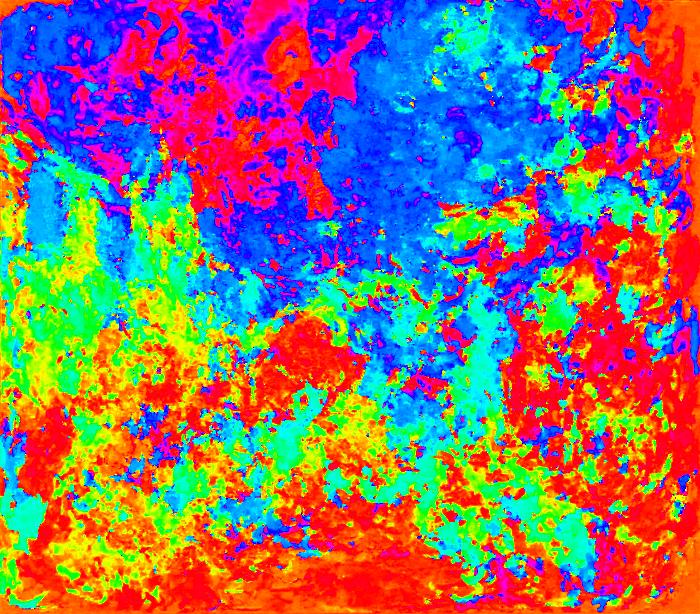}
  \caption{
    The hue field of the painting in Fig.~\ref{fig:rose_garden}, created by
    uniformly setting the brightness and saturation to one.
  }
\end{figure}

\begin{figure}
  \centering
  \includegraphics[width=0.26\textwidth]{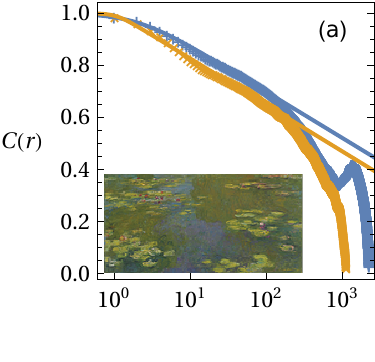}\hspace{-1em}
  \includegraphics[width=0.26\textwidth]{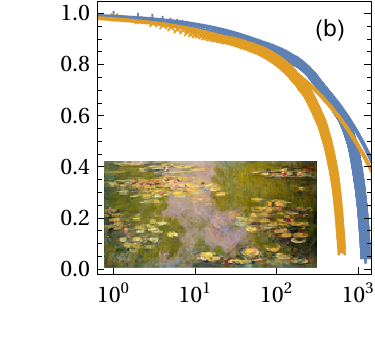}\hspace{-1em}
  \includegraphics[width=0.26\textwidth]{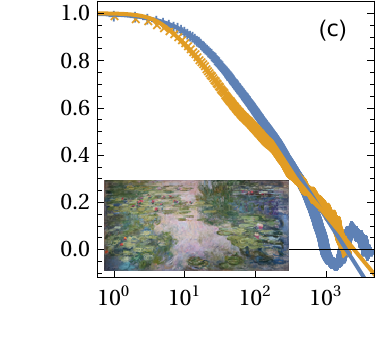}\hspace{-1em}
  \includegraphics[width=0.26\textwidth]{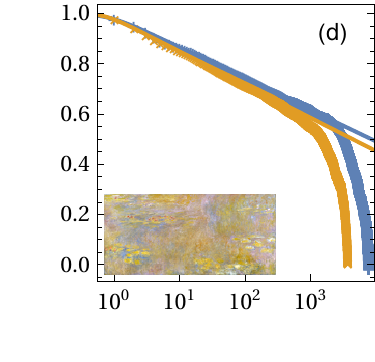}
  \vspace{-0.75em}

  \includegraphics[width=0.26\textwidth]{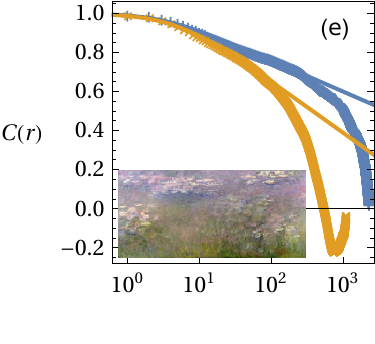}\hspace{-1em}
  \includegraphics[width=0.26\textwidth]{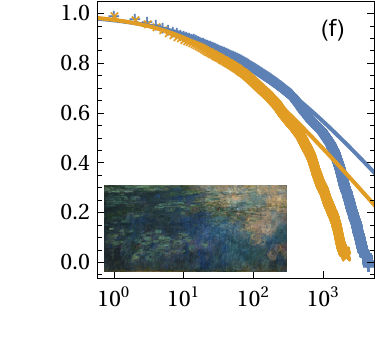}\hspace{-1em}
  \includegraphics[width=0.26\textwidth]{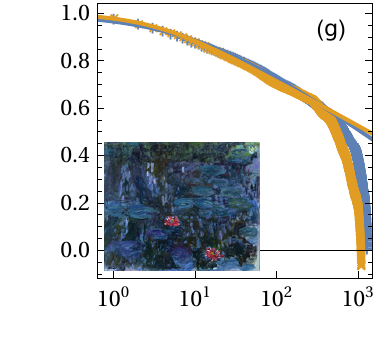}\hspace{-1em}
  \includegraphics[width=0.26\textwidth]{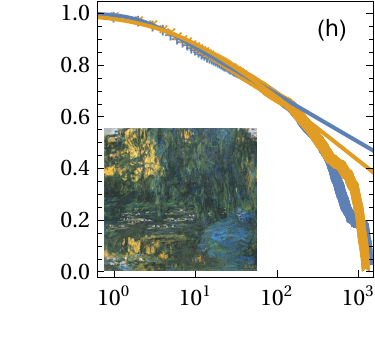}
  \vspace{-0.75em}

  \includegraphics[width=0.26\textwidth]{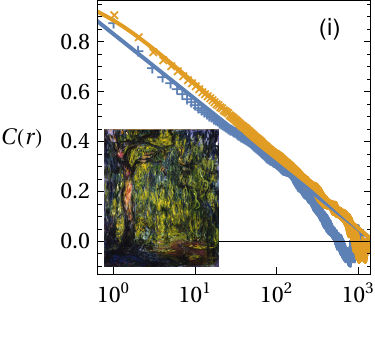}\hspace{-1em}
  \includegraphics[width=0.26\textwidth]{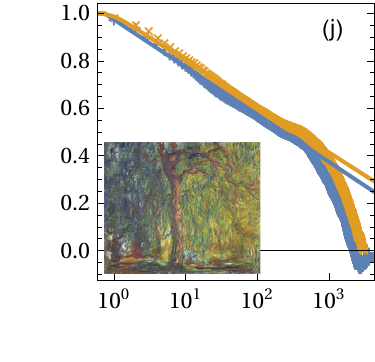}\hspace{-1em}
  \includegraphics[width=0.26\textwidth]{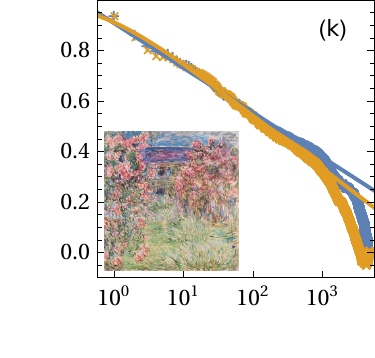}\hspace{-1em}
  \includegraphics[width=0.26\textwidth]{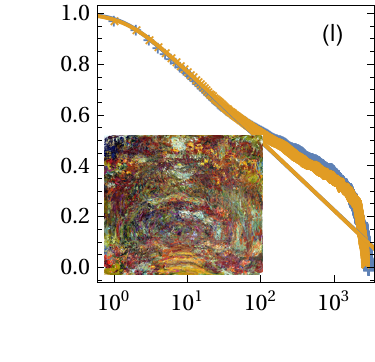}
  \vspace{-0.75em}

  \includegraphics[width=0.26\textwidth]{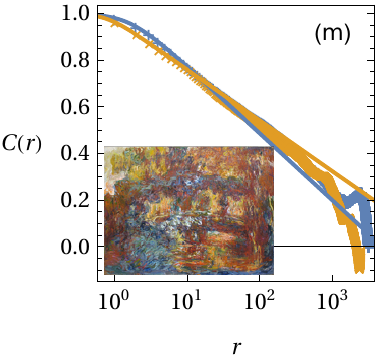}\hspace{-1em}
  \includegraphics[width=0.26\textwidth]{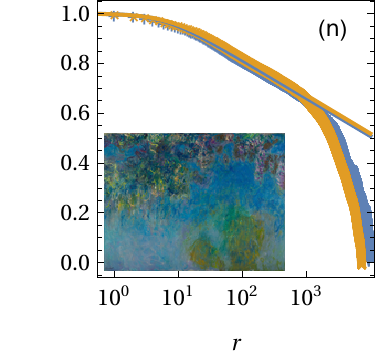}\hspace{-1em}
  \includegraphics[width=0.26\textwidth]{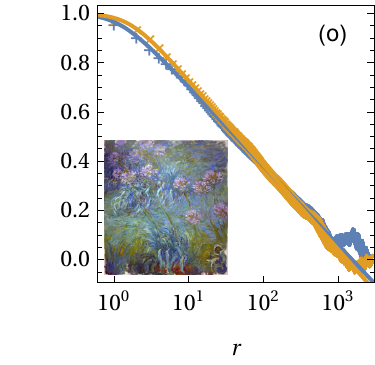}\hspace{-1em}
  \includegraphics[width=0.26\textwidth]{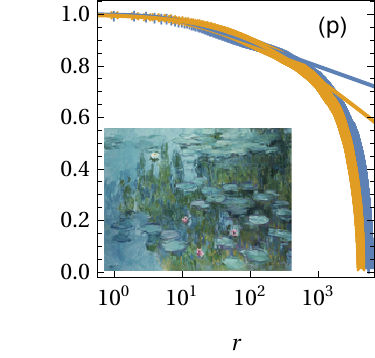}

  \caption{
    Correlation functions in the horizontal (blue) and vertical (orange)
    directions, along with fits of cutoff logarithms, for a variety of Monet
    paintings. The paintings are, from left to right and top to bottom,
    \textbf{(a)} \href{https://en.wikipedia.org/wiki/File:Le_bassin_aux_nymph\%C3\%A9as_-_Claude_Monet.jpg}{\textit{Le bassin aux nymphéas} (1919)},
    \textbf{(b)} \href{https://en.wikipedia.org/wiki/File:WLA_metmuseum_Water_Lilies_by_Claude_Monet.jpg}{\textit{Water lilies} (1919)},
    \textbf{(c)} \href{https://en.wikipedia.org/wiki/File:Claude_Monet_-_Water_Lilies,_1917-1919.JPG}{\textit{Water lilies} (1917--1919)},
    \textbf{(d)} \href{https://en.wikipedia.org/wiki/File:Claude_Monet_044.jpg}{\textit{Water lilies} (1920)},
    \textbf{(e)} \href{https://en.wikipedia.org/wiki/File:Claude_Monet,_Water_Lilies,_ca._1915-1926.jpg}{\textit{Water lilies} (c.~1915--1926)},
    \textbf{(f)} \href{https://en.wikipedia.org/wiki/File:Claude_Monet_-_Reflections_of_Clouds_on_the_Water-Lily_Pond.jpg}{\textit{Reflection of clouds on the water-lily pond} (c.~1920)},
    \textbf{(g)} \href{https://en.wikipedia.org/wiki/File:Nymph\%C3\%A9as_reflets_de_saule_1916-19.jpg}{\textit{Water lilies and reflections of a willow} (1916--1919)},
    \textbf{(h)} \href{https://en.wikipedia.org/wiki/File:Claude_Monet,_Water-Lily_Pond_and_Weeping_Willow.JPG}{\textit{Water-lily pond and weeping willow} (1916--1919)},
    \textbf{(i)} \href{https://en.wikipedia.org/wiki/File:Claude_Monet,_Weeping_Willow.JPG}{\textit{Weeping willow} (1918)},
    \textbf{(j)} \href{https://en.wikipedia.org/wiki/File:Claude_Monet_Weeping_Willow.jpg}{\textit{Weeping willow} (1918--1919)},
    \textbf{(k)} \href{https://en.wikipedia.org/wiki/File:Monet_-_Das_Haus_in_den_Rosen.jpeg}{\textit{House among the roses} (1917--1919)},
    \textbf{(l)} \href{https://en.wikipedia.org/wiki/File:Monet-_Der_Rosenweg_in_Giverny.jpeg}{\textit{The rose walk, Giverny} (1920--1922)},
    \textbf{(m)} \href{https://en.wikipedia.org/wiki/File:1920-22_Claude_Monet_The_Japanese_Footbridge_MOMA_NY_anagoria.JPG}{\textit{The Japanese footbridge} (1920--1922)},
    \textbf{(n)} \href{https://en.wikipedia.org/wiki/File:Claude_Monet_-_Wisteria_-_Google_Art_Project.jpg}{\textit{Wisteria} (1920--1925)},
    \textbf{(o)} \href{https://en.wikipedia.org/wiki/File:1914-26_Claude_Monet_Agapanthus_MOMA_NY_anagoria.JPG}{\textit{Agopanthus} (1914--1926)}, and
    \textbf{(p)} \href{https://en.wikipedia.org/wiki/File:Nympheas_71293_3.jpg}{\textit{Water lilies} (c.~1915)}.
  } \label{fig:examples}
\end{figure}

In order to further investigate this, we chose a collection of Monet paintings in
similar style, all painted in the late part of the artist's life. These were
all analysed as described above, with correlation functions plotted in
Fig.~\ref{fig:examples} and fit to logarithms with a short-distance cutoff.
Most of the paintings have some scale at which the logarithmic correlations
describe the behavior reasonably well, but the quality of the logarithmic
description varies substantially from painting to painting.

\begin{figure}
  \includegraphics[width=0.51\textwidth]{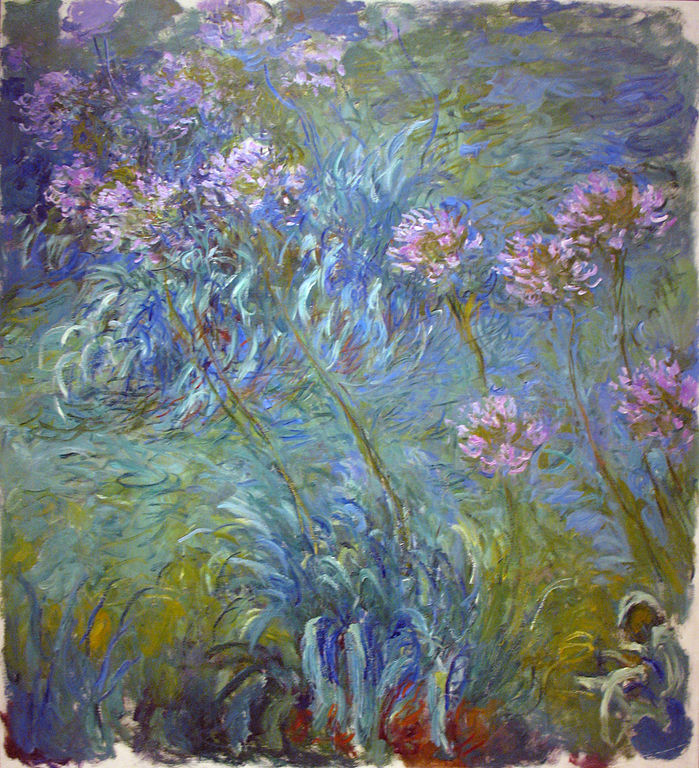}
  \hfill
  \includegraphics[width=0.47\textwidth]{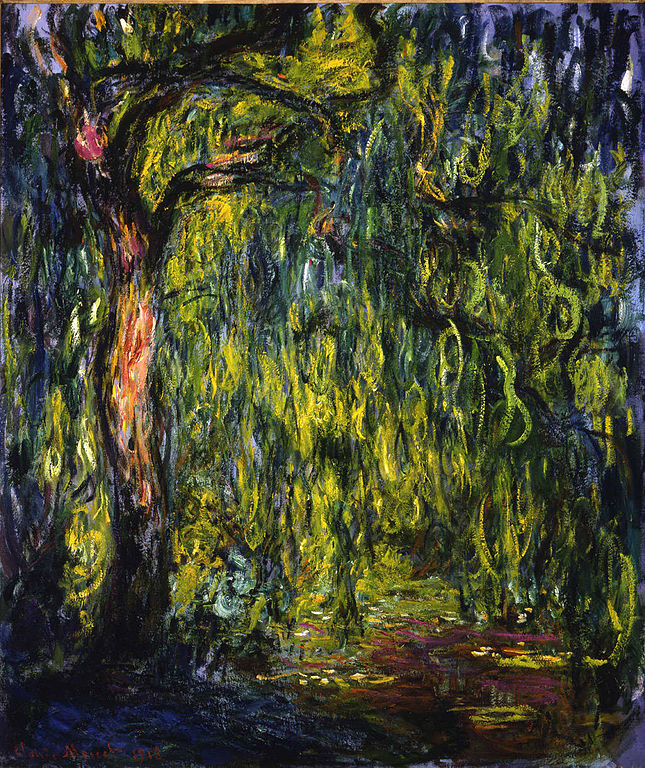}

  \vspace{0.5em}

  \includegraphics[width=0.4\textwidth]{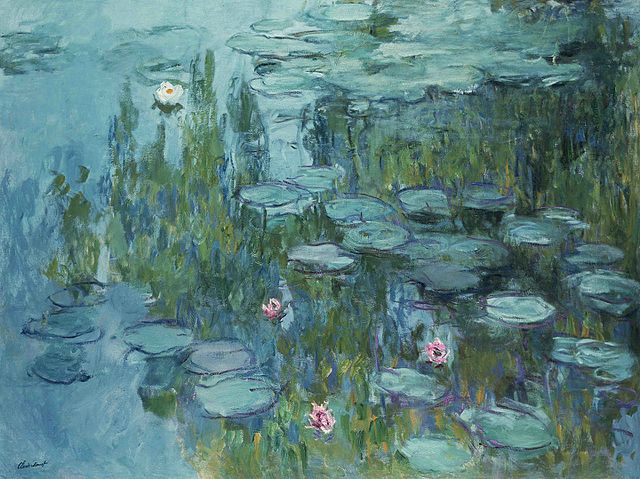}
  \hfill
  \includegraphics[width=0.59\textwidth]{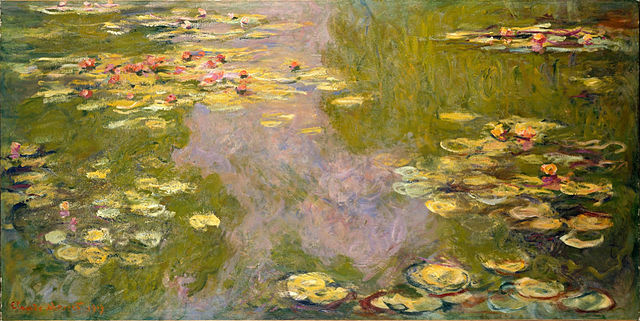}

  \caption{
    Examples of the Monet paintings in Fig.~\ref{fig:examples} the strongest
    and weakest log correlation. \textbf{Top:} Paintings with the strongest log
    correlation:
    \href{https://en.wikipedia.org/wiki/File:1914-26_Claude_Monet_Agapanthus_MOMA_NY_anagoria.JPG}{\textit{Agopanthus} (1914--1926)}
    and
    \href{https://en.wikipedia.org/wiki/File:Claude_Monet,_Weeping_Willow.JPG}{\textit{Weeping willow} (1918)}.
    \textbf{Bottom:} Paintings with the weakest log correlation:
    \href{https://en.wikipedia.org/wiki/File:Nympheas_71293_3.jpg}{\textit{Water lilies} (c.~1915)}
    and
    \href{https://en.wikipedia.org/wiki/File:WLA_metmuseum_Water_Lilies_by_Claude_Monet.jpg}{\textit{Water lilies} (1919)}.
  } \label{fig:comparison}
\end{figure}

Fig.~\ref{fig:comparison} compares paintings with the best and worst
descriptions by log correlations. Certain qualitative features can be
identified by eye: the paintings with stronger log correlations seem to have
more color variations at short and medium length scales, while those with weak
log correlations have large regions of slow color variation with sharper
boundaries. Paintings with stronger log correlations also appear more abstract.

\begin{figure}
  \centering
  \includegraphics[width=0.51\textwidth]{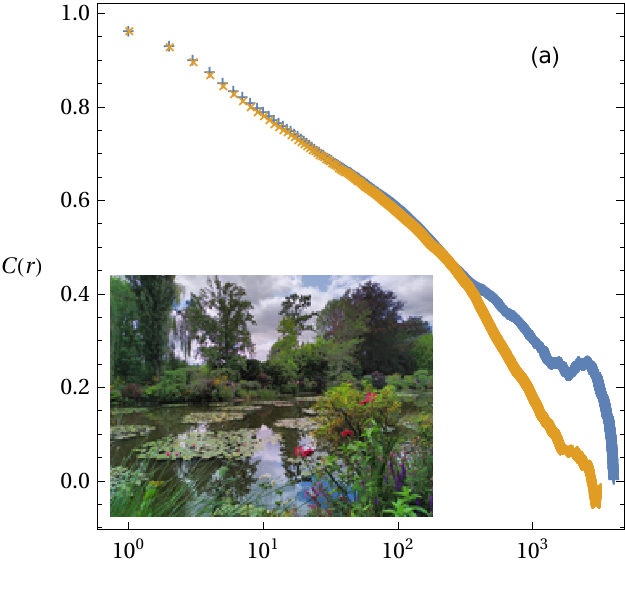}
  \hspace{-1.5em}
  \includegraphics[width=0.51\textwidth]{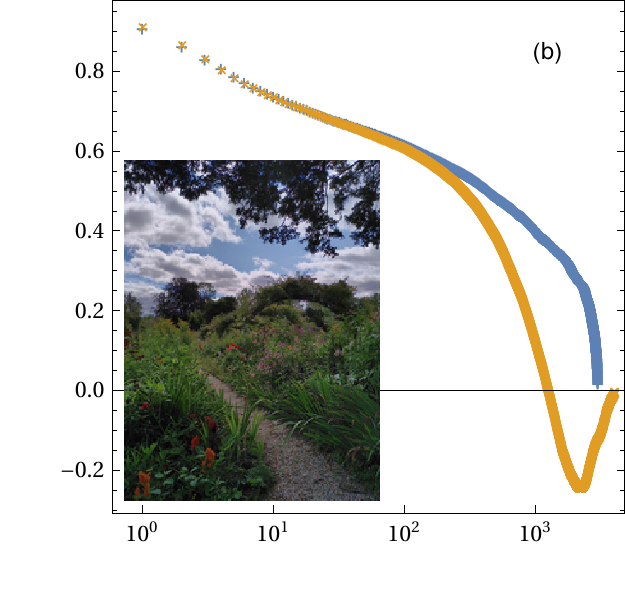}

  \vspace{-0.75em}

  \includegraphics[width=0.51\textwidth]{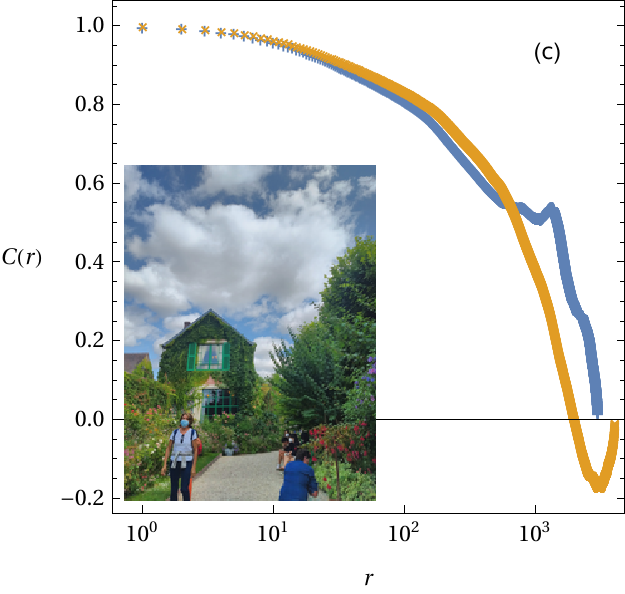}
  \hspace{-1.5em}
  \includegraphics[width=0.51\textwidth]{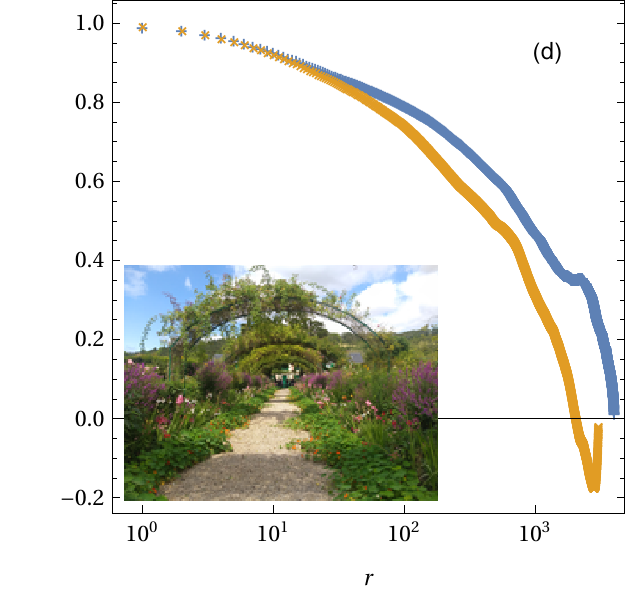}

  \caption{
    Correlation functions in the horizontal (blue) and vertical (orange)
    directions for several photographs of Claude Monet's gardens in Giverny.
    These photos were taken by the author or his wife on August 19th 2021.
    \textbf{(a)} \href{https://kent-dobias.com/imgs/IMG_20210819_123549Z_HDR.jpg}{The lily pond from the south bank}.
    \textbf{(b)} \href{https://kent-dobias.com/imgs/IMG_20210819_130041Z_HDR.jpg}{The gardens from the artist's house}.
    \textbf{(c)} \href{https://kent-dobias.com/imgs/IMG_20210819_153240_HDR.jpg}{The artist's house from the rose garden}.
    \textbf{(d)} \href{https://kent-dobias.com/imgs/IMG_20210819_145505_1.jpg}{The artist's house through the trestles}.
  } \label{fig:natural}
\end{figure}

It's known that photographs of natural scenes have fluctuations characteristic
of logarithmic correlations \cite{Ruderman_1994_The, Ruderman_1994_Statistics,
Stephens_2013_Statistical}. Could it be that Claude Monet has just faithfully
reproduced these actually occurring statistics in the natural scenes he
depicted? To investigate this, we preformed the same analysis with several
photographs of Monet's gardens in Giverny taken by the author or his wife,
shown in Fig.~\ref{fig:natural}. Like other natural images, these photos
exhibit varying levels of log correlations, or at least their signature. This
suggests that Monet is capturing the real correlations in the scenes he
depicts, and in his more abstract work amplifies them dramatically.

Despite the common belief that beauty is subjective, attempts have been made to
quantify the beauty of art. In work by Lakhal \emph{et al}.\
\cite{Lakhal_2020_Beauty}, the authors propose that structural complexity could
be such a measure. They had people compare randomly generated images with
particular spatial correlations; log correlations correspond to their parameter
$\alpha=-1$. They found that the people surveyed found log correlated images to
be among the most attractive, perhaps because of their statistical connection
to natural images. Therefore, one might reason that much of wildly-appreciated
art would contain log correlations. What makes some of these Monet paintings so
striking is how well-described their correlations are by logarithms alone.

Many of the Monet paintings with strong log correlations seem to have been done
during the period 1914--1925, a time when the artist suffered from cataracts
affecting his vision \cite{Gruener_2015_The, Marmor_2006_Ophthalmology}. At the
beginning of this period, he reported that ``colours no longer had the same
intensity for me,'' ``reds had begun to look muddy,'' and ``my painting was
getting more and more darkened.'' His work took on a broader style with more
intense colors, epitomized by Fig.~\ref{fig:rose_garden}. Perhaps Monet's
cataracts affected his painting by leading him to emphasize the abstract
(log-correlated) beauty in the strokes of paint over the literal (less
log-correlated) beauty of the scene they were meant to depict. However, after
1923--1925 when his cataracts were surgically treated, Monet destroyed many of
his paintings from the preceding period and returned to an earlier style. This
suggests that the log correlations alone were not enough to meet Monet's own
standard of beauty.

\begin{figure}
  \centering
  \includegraphics[width=\textwidth]{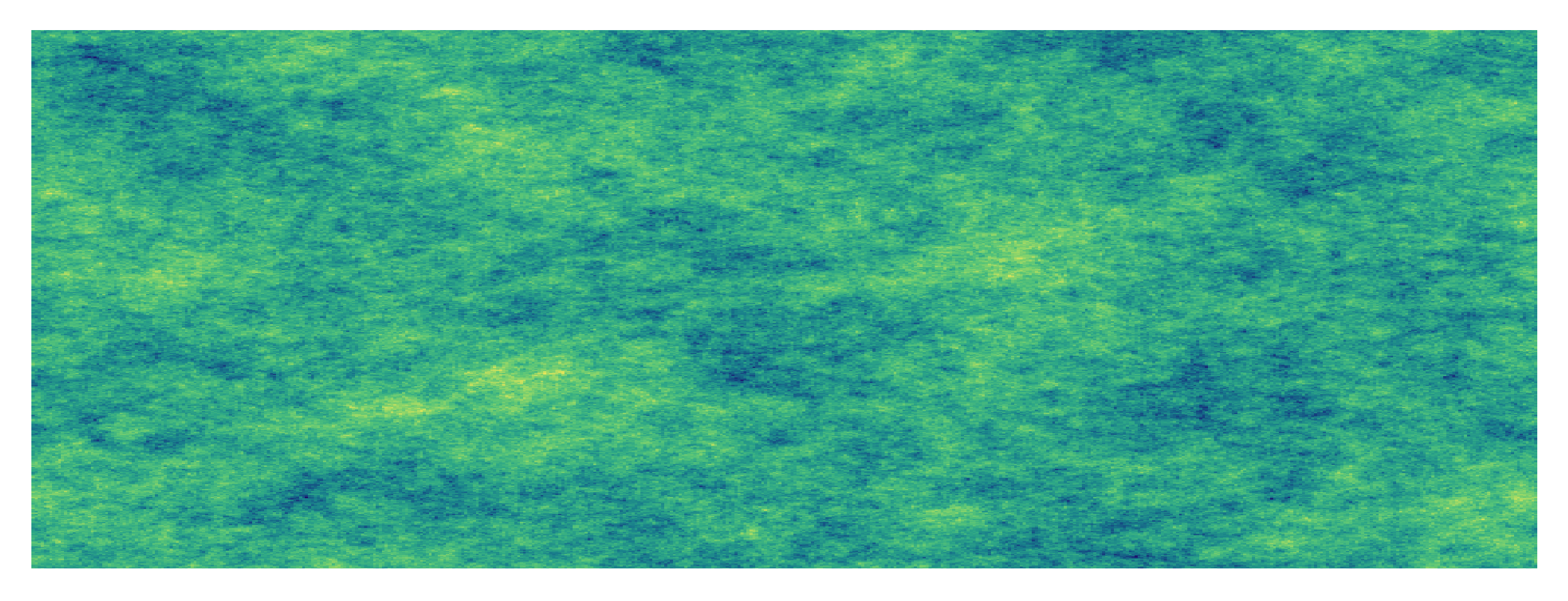}

  \caption{
    The same Gaussian free field as in Fig.~\ref{fig:gff}, but scaled
    anisotropically and colored using the \textsl{Mathematica}
    \texttt{BlueGreenYellow} color scheme.
  } \label{fig:fake}
\end{figure}

The presence of logarithmic correlations within Monet paintings suggests a
method for producing new Monet-like graphics. First, a log-correlated
configuration of some system is produced by standard means. Then, the field is
colored using a suitable palette, and perhaps stretched or squished to produce
effective anisotropy between horizontal and vertical directions. A first
attempt to produce such an image is found in Fig.~\ref{fig:fake}, which the
authors feel has a passing resemblance to some of Monet's work. Incorporating
the appropriate brightness and saturation fluctuations would likely improve the
resemblance further.

\paragraph{Acknowledgements}
The authors would like to thank Yan Fyodorov for his many invaluable
contributions to physics and mathematics (especially that of log-correlated
random fields) and wish him a happy birthday. We would also like to thank
Valentina Ros for the inspiration to investigate this subject and for help with
the literature. All images of paintings in this article were sourced from
Wikipedia.

\paragraph{Funding information}
JK-D is supported by the Simons Foundation Grant No.~454943.

\printbibliography

\end{document}